\documentclass[sigconf, 9pt]{acmart}
\AtBeginDocument{%
  }

\usepackage{tikz}
\usetikzlibrary{shapes.geometric, arrows, positioning, fit, backgrounds}
\usepackage{amsmath}
\usepackage{booktabs}
\usepackage{multirow}
\usepackage{graphicx}
\usepackage{amsmath}
\usepackage{microtype}
\usepackage{url}
\usepackage{xcolor}
\usepackage{subcaption}
\usepackage{colortbl}
\usepackage[table]{xcolor}
\usepackage{makecell}
\newcommand{\prism}{\textsc{Prism}}
\newcommand{\voltage}{\textsc{Voltage}}
\newcommand{\sm}{Segment Means}
\copyrightyear{2026}
\acmYear{2026}
\setcopyright{cc}
\setcctype{by}
\acmConference[MobiSys Workshop '26]{The 24th Annual International Conference on Mobile Systems, Applications and Services}{June 21--25, 2026}{Cambridge, United Kingdom}
\acmBooktitle{The 24th Annual International Conference on Mobile Systems, Applications and Services (MobiSys Workshop '26), June 21--25, 2026, Cambridge, United Kingdom}
\acmDOI{10.1145/3812836.3814999}
\acmISBN{979-8-4007-2712-2/2026/06}

\begin{document}
%\title{Profiling-Driven Adaptive Distributed Transformer Inference on a Jetson Cluster}
 \title{Profiling-Driven Adaptive Distributed Transformer Inference on Embedded Edge Deployment}
\author{Muhammad Azlan Qazi}
% \authornote{Authors contributed equally to this research.}
\affiliation{%
  \institution{Aarhus University}
  \city{Aarhus}
  \country{Denmark}}
\email{maq@ece.au.dk}
\author{Alexandros Iosifidis}
\affiliation{%
  \institution{Tampere University}
  \city{Tampere}
  \country{Finland}}
\email{alexandros.iosifidis@tuni.fi}
\author{Qi Zhang}
\affiliation{%
  \institution{Aarhus University}
  \city{Aarhus}
  \country{Denmark}}
\email{qz@ece.au.dk}

% \author{Muhammad Azlan Qazi$^1$, Alexandros Iosifidis$^{2}$, Qi Zhang$^3$ }
% \affiliation{%
% \institution{$^{1,3}$Aarhus University, Denmark}
% }
% \affiliation{%
% \institution{$^2$Tampere University, Finland}
% }
% \affiliation{%
% \institution{maq@ece.au.dk, alexandros.iosifidis@tuni.fi, qz@ece.au.dk}
% }

%% ==============================================================
% \begin{abstract}
% Distributing Transformer inference across embedded edge devices
% can mitigate individual memory and computational constraints,
% yet its real-world efficacy remains obscured by simulations
% that overlook hardware-specific communication overheads.
% We present a hardware prototype study of \prism{} and \voltage{},
% evaluated on NVIDIA Jetson Orin Nano modules connected via WiFi.
% Our key finding is a critical hardware bottleneck: Jetson
% boards integrate the GPU directly with the memory controller,
% eliminating the PCIe/NVLink pathway \texttt{NCCL} requiresT
% and forcing use of \texttt{GLOO}, where every communication data stages through CPU memory; an overhead scaling with
% communication data volume that renders full-tensor exchange slower than single-device inference across batch sizes.
% We use a lightweight offline profiling mechanism that
% characterizes per-sample latency and energy across batch sizes
% and network conditions, enabling \prism{} to \emph{adaptively}
% select the optimal execution mode at runtime.
% Measurements confirm \prism{}'s Segment Means compression
% yields 65\%--77\% latency reduction and 34\%--52\% energy
% savings over full-tensor exchange.
% \end{abstract}

\begin{abstract}
Distributing Transformer inference across embedded edge devices can alleviate individual memory and compute constraints, yet practical benefits on real hardware remain unclear: prior work relies largely on simulations that overlook hardware-specific communication overheads. We present a hardware prototype study on NVIDIA Jetson Orin Nano devices connected over WiFi. Our key finding is that the dominant bottleneck is not just network bandwidth but also the CPU–GPU staging during communication. Because Jetson's integrated GPU architecture lacks the PCIe/NVLink pathway that \texttt{NCCL} requires, all inter-device data communication should be routed through \texttt{GLOO} and staged in CPU memory; an overhead that scales with communication data volume and makes full-tensor exchange slower than single-device inference across the batch sizes for medium sized models such as ViT. We therefore evaluate \prism{} by combining Segment Means compression with lightweight offline profiling to adaptively select between local and distributed execution at runtime. Experiments show that this strategy reduces latency by 65\%–77\% and energy consumption by 34\%–52\% relative to full-tensor exchange in static distributed execution setup, demonstrating that profiling-driven adaptation is essential for practical distributed Transformer inference on embedded hardware.
\end{abstract}

\begin{CCSXML}
<ccs2012>
   <concept>
       <concept_id>10010147.10010178.10010219.10010223</concept_id>
       <concept_desc>Computing methodologies~Cooperation and coordination</concept_desc>
       <concept_significance>500</concept_significance>
       </concept>
   <concept>
       <concept_id>10010583.10010717</concept_id>
       <concept_desc>Hardware~Hardware validation</concept_desc>
       <concept_significance>300</concept_significance>
       </concept>
   <concept>
       <concept_id>10003033.10003079</concept_id>
       <concept_desc>Networks~Network performance evaluation</concept_desc>
       <concept_significance>100</concept_significance>
       </concept>
 </ccs2012>
\end{CCSXML}

\ccsdesc[500]{Computing methodologies~Cooperation and coordination}
\ccsdesc[300]{Hardware~Hardware validation}
\ccsdesc[100]{Networks~Network performance evaluation}

\keywords{distributed inference, edge computing, Transformer,
  NVIDIA Jetson, adaptive inference, inter-device communication, embedded systems, performance profiling, system-on-chip (SoC)}
\maketitle

%% ==============================================================
\section{Introduction}
\label{sec:introduction}

Transformer models used in vision and language tasks are
increasingly deployed on embedded AI systems.
On-device image classification, local NLP inference, and
real-time object detection all increasingly demand Transformer
models that run without cloud connectivity, on hardware such as
the NVIDIA Jetson~Orin~Nano: capable, affordable, but
constrained in memory and compute.
One natural strategy is to pool the resources of several such
devices through distributed inference.
Position-wise partitioning~\cite{10631032} is a promising
approach: the input token sequence is split along the sequence
dimension, each device processes its own partition
independently, and a single collective communication operation per transformer block
synchronizes intermediate features.
Simulation studies show strong throughput gains compared to Tensor Parallelism (TP)~\cite{Shoeybi2019MegatronLMTM}, but
translating these to real embedded hardware introduces challenges that prior work has not addressed.
\begin{figure}
\centering
\includegraphics[width=\columnwidth]{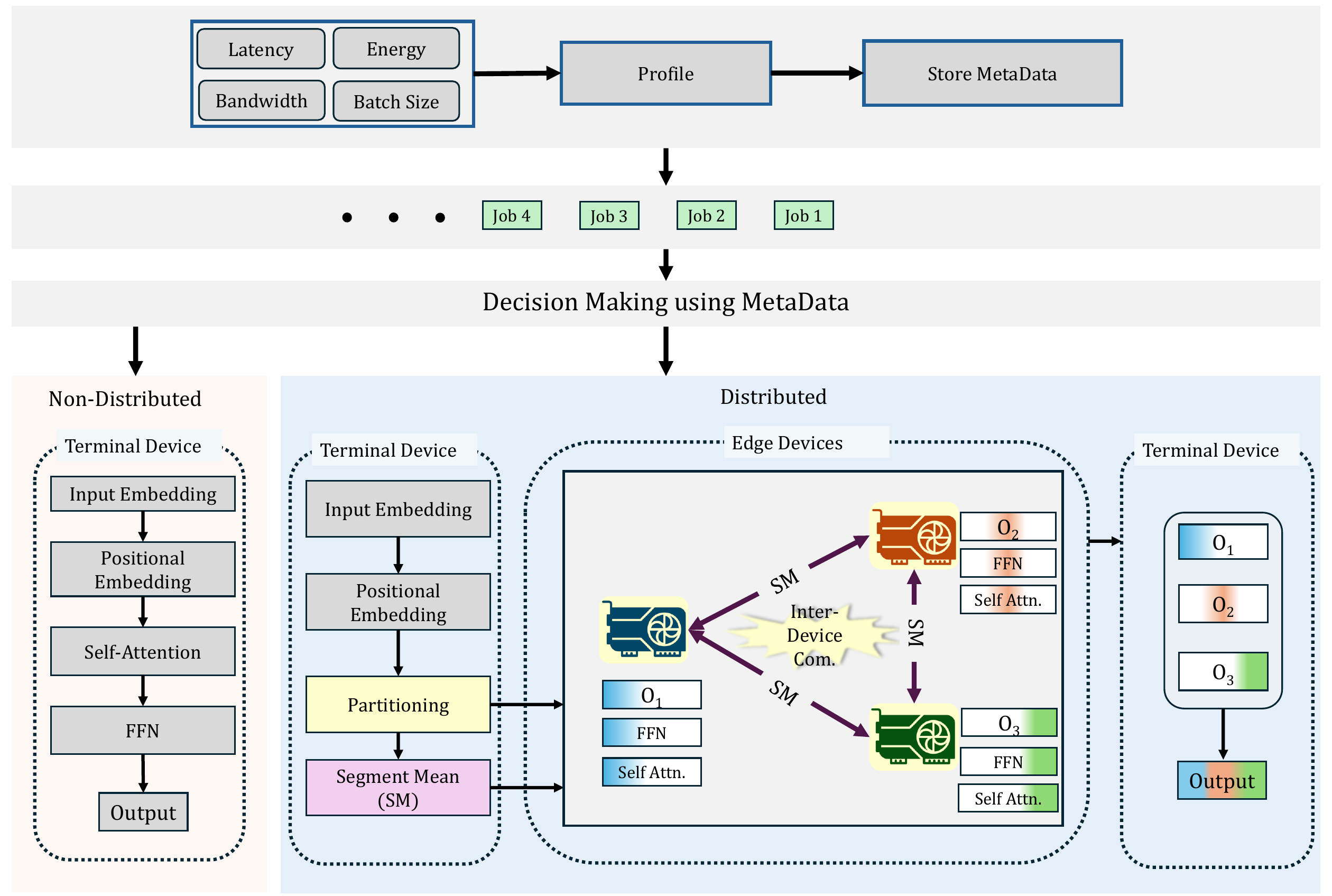}
\caption{Overview of \prism{}, showing offline profiling for adaptive local or distributed inference with compact \sm{} communication.}
\label{fig:arch}
\end{figure}

\textbf{The GLOO staging problem.}
Jetson boards come with the integrated GPU, i.e.
there are no PCIe-attached or NVLink-connected GPUs.
The NVIDIA Collective Communications Library (\texttt{NCCL}) backend, 
which performs direct GPU-to-GPU transfers in server deployments, is therefore
unavailable.
The only feasible option is \texttt{GLOO}~\cite{gloosite}, which requires
every communicated tensor to be staged through host CPU memory
before transmission and copied back to GPU memory on receipt.
This staging overhead scales with communication volume and
dominates latency in full-tensor exchange.

\textbf{Batch-size sensitivity.}
Edge applications operating at small batch sizes
for mid-sized models (such as ViT), where the fixed cost
of communication and staging may outweigh the compute savings
from distributing the workload.

\textbf{Network variability.}
% Network bandwidth fluctuates in practice. The benefit of distributed inference depends on whether network latency or staging latency dominates, a question that can only be answered by profiling on real hardware.
Network bandwidth fluctuates in practice. The benefit of distributed inference depends on whether network or staging latency dominates, which can only be determined through profiling on real hardware.

We address these challenges on \prism{}~\cite{qazi2025prism}, and report its hardware prototype study on a
Jetson~Orin~Nano boards. Fig.~\ref{fig:arch} shows the distributed inference setup of \prism{}, where input tokens are partitioned across edge devices and synchronized using compact \sm{} intermediate features. An offline profiling phase collects performance metadata, which is queried at runtime to adaptively switch between local and distributed inference based on batch size and network conditions. %The distributed setup is presented in Fig.~\ref{fig:arch}.

While \prism{}~\cite{qazi2025prism} introduces Segment Means
communication and evaluates it primarily in simulation,
its behavior on real embedded hardware remains unknown.
This paper therefore uses \prism{} as a case study to
characterize how distributed Transformer inference behaves
under the hardware constraints of integrated-GPU edge devices.
  Our contributions are:
\begin{itemize}
  \item \textbf{Develop a working prototype.}
  % \item \textbf{Hardware testbed.}
   We develop a working prototype using Jetson Orin Nano instrumented for 
   latency breakdown and per‑sample energy measurement. Using this prototype, we quantify
   the CPU–GPU staging overhead imposed by the \texttt{GLOO}~\cite{gloosite} communication 
   backend and show that it renders full‑tensor exchange (\voltage{}~\cite{10631032}) 
   counterproductive at all tested batch sizes.
\item \textbf{Profiling-driven adaptive inference policy.}
An offline profiling sweep across batch sizes and network conditions builds a performance map that enables runtime selection between single-device and distributed execution. We identify the batch-size crossover point below which local execution is preferred.   
  % \item \textbf{Profiling-driven adaptive inference policy.}
  %   An offline profiling mechanism sweeps batch size and
  %   bandwidth conditions, stores a performance map, and
  %   enables runtime selection between single-device and
  %   distributed execution.
  %   We characterize the batch-size crossover point below
  %   which local execution is preferred.
  \item \textbf{Bandwidth sensitivity measurements.}
    Hardware experiments at bandwidths from 200 to 900~Mbps
    reveal that \prism{}'s compact \sm{}
    communication maintains low latency across all tested
    conditions, while \voltage{} degrades severely at
    lower bandwidth.
\end{itemize}
%% ==============================================================
\section{Background and Related Work}
\label{sec:background}
% \subsection{Distributed Transformer Inference}
\label{sec:distributed_inference}
Parallelism strategies for distributed Transformer inference
span a range of communication--computation trade-offs.
Model parallelism~\cite{NIPS2012_6aca9700} splits the model into sub-graphs
(groups of layers) assigned to compute nodes, passing activations sequentially
and causing pipeline bubbles (device idle time) that are especially costly at
small batch sizes.
Pipeline parallelism~\cite{10.5555/3454287.3454297} mitigates
idle time with micro-batches but provides minimal benefit for edge computation
because devices still experience idle time.
Tensor parallelism~\cite{Shoeybi2019MegatronLMTM} splits
individual weight matrices and requires two
\texttt{AllReduce} operations per Transformer block;
prior work found this demands at least 1000~Mbps to
outperform single-device inference with
six participating devices~\cite{10631032}.

\voltage{}~\cite{10631032} uses position-wise
partitioning: the input sequence is split across devices and a single
\texttt{AllGather} per block collects intermediate features,
reducing data communication per transformer block to $(P{-}1)ND/P$, where $P$ is the number of data partitions or available edge devices, $N$ is the number of input tokens or sequences, and $D$ is the dimensionality of the attention features.
While efficient relative to tensor parallelism, it still
transmits full intermediate features, which causes increased in staging overhead on Jetson.

\prism{}~\cite{qazi2025prism} extends position-wise
partitioning by replacing full intermediate feature with compact
\sm{}: column-wise averages of
non-overlapping token segments.
Communication reduces to $(P{-}1)LD$ elements, where
$L \ll N/P$ is controlled by the compression rate
$\mathrm{CR} = N/(L \cdot P)$.
A scaling-aware softmax reformulation further eliminates
redundant Key/Value recomputation across devices, reducing per-device
GFLOPs by up to 50.11\% with two devices.
The present paper focuses on prototype design, profiling,
and hardware evaluation. %; simulation results and accuracy
%analysis can be found in~\cite{qazi2025prism}.

% \subsection{Benchmarking Inference on Embedded Hardware}
% \label{sec:benchmarking}
% Studies of neural accelerators on mobile platforms have
% found that specialised hardware does not always outperform
% conventional mobile GPUs and DSPs, and that CPU--GPU data
% movement is a critical, underappreciated bottleneck on
% constrained SoCs~\cite{cao2021mobile}.
% Benchmarking of video object detection on Jetson boards
% showed that adaptive policies responding to runtime resource
% availability consistently outperform static
% configurations~\cite{lee2021benchmarking}.
% Our work extends this principle to distributed Transformer
% inference, where the adaptive question is not which
% configuration to use but whether to distribute at all.

%% ==============================================================
\section{System Overview}
\label{sec:system}
\subsection{Segment Means Communication}
\label{sec:segmean}
\prism{} follows a master--worker architecture as illustrated in Fig.~\ref{fig:arch}.
A terminal device partitions the input
$\mathbf{X} \in \mathbb{R}^{N \times D}$ into $P$ equal
parts $\{\mathbf{X}_1, \ldots, \mathbf{X}_P\}$ along the
sequence dimension.
For each partition $\mathbf{X}_p \in \mathbb{R}^{N_p \times D}$,
it computes $L$ \sm{} by dividing the partition into
$L$ equal segments and taking the column-wise mean of each:
\begin{equation}
  \mathbf{Z}_p = [\boldsymbol{\mu}_0;\,\boldsymbol{\mu}_1;
    \,\ldots;\,\boldsymbol{\mu}_{L-1}] \in \mathbb{R}^{L\times D}.
  \label{eq:segmean}
\end{equation}
Each device receives its own partition $\mathbf{X}_p$ and
the \sm{} $\mathbf{Z}_{p'}$ of every other device
$p' \neq p$, constructing the augmented representation:
\begin{equation}
  \hat{\mathbf{X}}_p =
    \mathrm{concat}(\mathbf{X}_p, \mathbf{Z}_j,
    \mathbf{Z}_k, \ldots),\quad
    (j,k,\ldots) \in \mathcal{I} \setminus \{p\}.
  \label{eq:augmented}
\end{equation}
Query is computed from $\mathbf{X}_p$; Key and Value use
$\hat{\mathbf{X}}_p$.
A scaling-aware softmax reformulation~\cite{qazi2025prism}
exploits the repetitive structure of Eq.~\eqref{eq:augmented}
to further reduce redundant computations.

The compression rate ($\mathrm{CR}$) controls
the trade-off between communication volume and accuracy.
Crucially for edge deployment, it also directly controls
the volume of data that must be staged through CPU memory,
making it the primary tuning knob for the adaptive policy.
The selected CR values $\{3.3, 4.95, 9.9\}$ correspond to
$L \in \{30, 20, 10\}$ segment means for $N{=}197$ tokens.
These represent low, medium, and high compression regimes
while keeping segment sizes integer, and allow us to study
how staging volume scales with communication reduction.
\subsection{The GLOO Staging Bottleneck}
\label{sec:gloo}

As the Jetson~Orin~Nano integrates its GPU with the shared memory
controller, there are no PCIe-attached or NVLink-connected
GPUs.
On Jetson-class integrated GPUs, \texttt{NCCL} falls back to host-staged TCP transfers due to the lack of PCIe/NVLink.
Consequently, it provides no practical benefit over \texttt{GLOO}, because all tensors must be copied GPU$\rightarrow$CPU before transmission and CPU$\rightarrow$GPU on receipt.
Thus, \texttt{NCCL} behaves effectively similar to \texttt{GLOO} in our deployment.

As a result, the only practical backend is \texttt{GLOO}, which operates
on CPU memory and introduces a two-step staging process
for every inter-device tensor:
% Jetson~Orin~Nano integrates its GPU with the shared memory
% controller; there are no PCIe-attached or NVLink-connected
% GPUs.
% % This rules out \texttt{NCCL}, which performs direct
% % GPU-to-GPU collective operations.
% \textcolor{red}{On Jetson-class integrated GPUs, \texttt{NCCL} falls back to host-staged TCP transfers due to the lack of PCIe/NVLink.
% Consequently, it provides no practical benefit over \texttt{GLOO}: all tensors must be copied GPU$\rightarrow$CPU before transmission and CPU$\rightarrow$GPU on receipt.
% Thus, \texttt{NCCL} behaves effectively the same as \texttt{GLOO} in our deployment.}
% The only available backend is \texttt{GLOO}, which operates
% on CPU memory and introduces a two-step staging process
% for every inter-device tensor:
\begin{enumerate}
  \item \textbf{Device$\to$Host:} The tensor is copied from
    GPU memory to pinned CPU memory before transmission.
  \item \textbf{Host$\to$Device:} The received tensor is
    copied from CPU memory back to GPU memory.
\end{enumerate}
Staging latency is proportional to tensor size and
largely independent of network bandwidth.
For \voltage{}~\cite{10631032}, which exchanges full
embedding vectors per layer, this staging completely
dominates the latency budget.
\prism{} reduces the staged data volume elements, directly shrinking the staging cost by the factor ($CR$) as the communication reduction.

\subsection{Profiling-Driven Adaptive Inference}
\label{sec:adaptive}

\begin{figure}[t]
  \centering
  %% Replace placeholder with your actual figure:
  \includegraphics[width=0.92\linewidth,height=0.28\textheight,
        keepaspectratio]{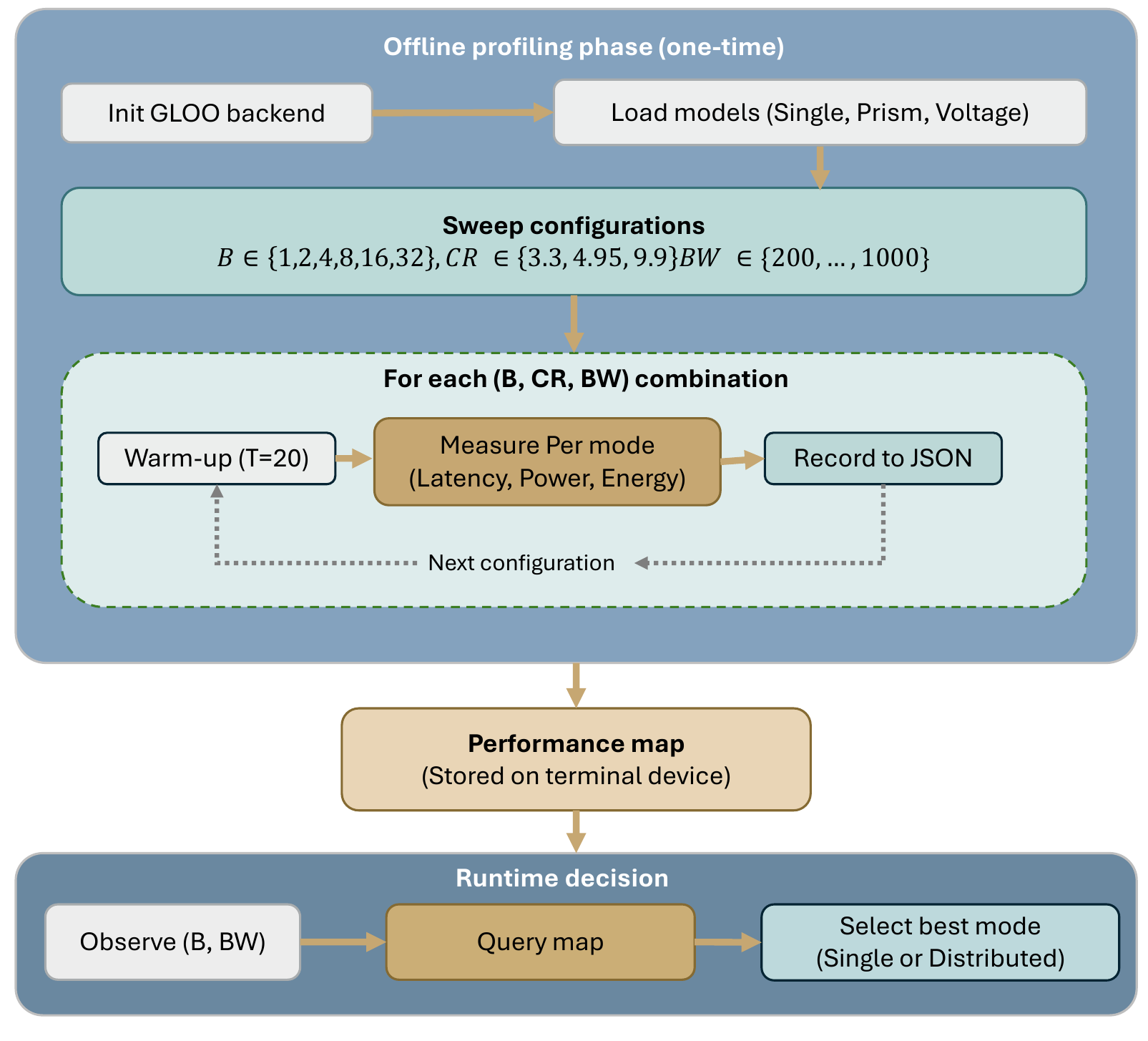}
  % \fbox{\rule{0pt}{3.0cm}\rule{0.88\linewidth}{0pt}}
  \caption{Profiling flow in \prism{} for adaptive runtime execution.}
  % \caption{Profiling flow in \prism{}.
  %   An offline sweep records per-sample latency and energy
  %   across batch sizes, compression rates, and bandwidth
  %   conditions.
  %   The resulting performance map is stored on the terminal
  %   device and queried at runtime to select single-device or
  %   distributed execution.}
  \label{fig:profiling_flow}
\end{figure}

A static execution mode is suboptimal because the optimal
choice depends on batch size, network bandwidth, and CR;
which are unknown at design time.
\prism{} uses a light offline profiling
phase (Fig.~\ref{fig:profiling_flow}) executed once
before deployment.

The profiler sweeps:
\begin{itemize}
  \item Batch size $B \in \{1, 2, 4, 8, 16, 32\}$,
  \item Compression rate $\mathrm{CR} \in \{3.3, 4.95, 9.9\}$,
  \item Bandwidth
    $BW \in \{200, 300, 400, 500, 600, 700, 800, 900\}$~Mbps
    (throttled via \texttt{tc netem} on the terminal device).
\end{itemize}
For each combination it records total latency (ms),
per-sample latency (ms), per-sample energy (J), and the
three-way breakdown among computation, communication,
and CPU--GPU I/O.
These measurements are stored as a performance map on the
terminal device, a lightweight JSON file requiring
negligible storage.

At runtime, when a batch of size $B$ arrives under
observed network conditions, the terminal device queries
the profiled map and selects the execution mode (non-distributed or
distributed with the best-performing CR) that minimizes
per-sample latency or energy, depending on the
application objective.
% The adaptive decision adds no additional communication latency and
% introduces only a constant-time table lookup at runtime.
The one-time profiling cost is approximately
$|{\mathcal{B}}| \times |{\mathcal{CR}}| \times
|{\mathcal{BW}}| \times T$ inference passes, where
we use $T{=}20$ warm-up runs per configuration.

%% ==============================================================
\section{Prototype Setup}
\label{sec:setup}

\begin{figure}[t]
  \centering
  \includegraphics[width=0.88\linewidth]{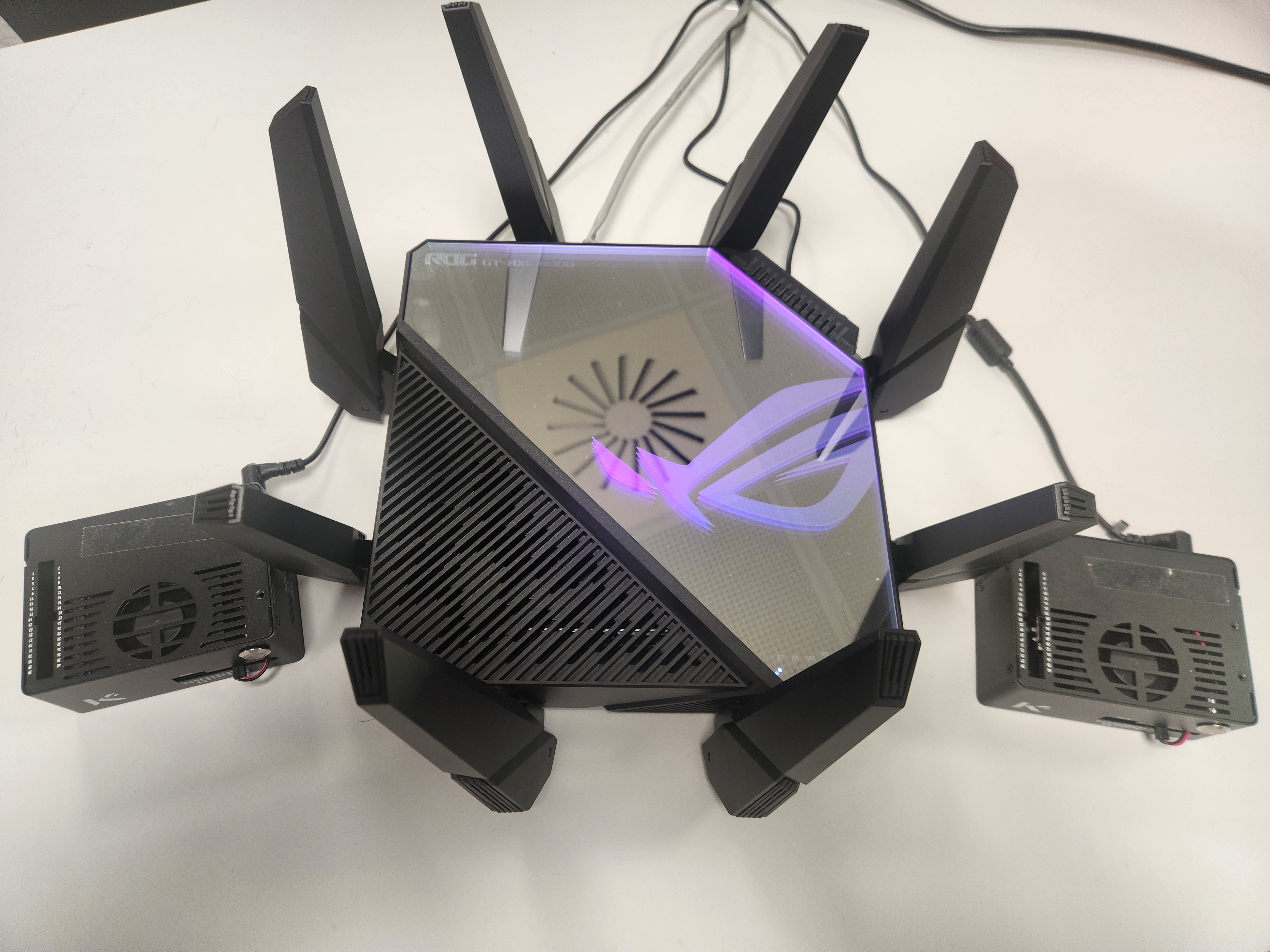}
  % \caption{Hardware prototype consisting of Jetson~Orin~Nano boards connected through a WiFi~6E router using Intel AX210 adapters.}
  \caption{Jetson~Orin~Nano hardware prototype connected over WiFi~6E.}
  \label{fig:prototype}
\end{figure}

\textbf{Hardware.}
Our prototype comprises NVIDIA Jetson~Orin~Nano development
kits connected to an ASUS ROG Rapture GT-AXE16000 WiFi~6E
router (Fig.~\ref{fig:prototype}).
Table~\ref{tab:hardware} summarizes the board specifications.
All boards are homogeneous Jetson Orin Nano units, and experiments use a two-board configuration. Although we evaluate two devices, staging overhead grows
linearly with the number of participants because each device
must stage additional received tensors.
Therefore, the crossover batch size where distributed
execution becomes beneficial would shift to larger values
as more nodes are added.

\begin{table}[t]
  \caption{NVIDIA Jetson Orin Nano hardware.}
  \label{tab:hardware}
  \centering
  \small
  \begin{tabular}{p{2.2cm} p{0.65\columnwidth}}
    \toprule
    \textbf{Component} & \textbf{Specification} \\
    \midrule
    SoC             & NVIDIA Orin (Ampere GPU + ARM Cortex) \\
    GPU cores       & 1024 CUDA cores (Ampere) \\
    Memory          & 8\,GB LPDDR5, unified CPU/GPU \\
    WiFi module     & Intel AX210, WiFi 6E (up to 1\,Gbps) \\
    Comm.\ backend  & \texttt{GLOO} (CPU-staged; \texttt{NCCL}
                      unavailable) \\
    Power budget    & 7--25\,W (configurable) \\
    \bottomrule
  \end{tabular}
\end{table}

\textbf{Software.}
We use PyTorch~2.x with \texttt{torch.distributed} and the
\texttt{GLOO} backend.
Communication and computation latencies are profiled with
the PyTorch Profiler (Fig.~\ref{fig:profiler_trace} shows flame graph produced by the tracer).
Per-sample energy is recorded by Jetson Stats
(\texttt{jtop}), which reads the on-board INA3221 power monitor.

\textbf{Bandwidth throttling.}
Network bandwidth is throttled using Linux \texttt{tc netem}
on the terminal device to simulate conditions from 200 to
900~Mbps.

\textbf{Model and workload.}
We evaluate \prism{} on ViT~\cite{dosovitskiy2021imageworth16x16words} with
CIFAR-10 inputs ($224{\times}224{\times}3$, $N{=}197$ tokens).
Batch size is swept from 1 to 32.
All distributed experiments compare \prism{} (CR~=~9.9
unless stated) against \voltage{}~\cite{10631032}
(full-tensor exchange) and single-device inference
as a lower-bound baseline. It is worth noting that, ViT is used as a representative Transformer workload because
its communication pattern mirrors that of encoder and decoder
Transformer blocks used in vision, language, and multimodal
foundation models.

\section{Evaluation}
\label{sec:evaluation}

We evaluate \prism{} on two Jetson~Orin~Nano boards using ViT with CIFAR-10.
Section~\ref{sec:latency} analyses the latency breakdown and identifies the staging bottleneck.
Sections~\ref{sec:per_sample} examine per-sample efficiency.
Section~\ref{sec:qualitative} validates prediction correctness, and Section~\ref{sec:profiling_instrumentation} describes the profiling instrumentation.
Section~\ref{sec:insights} concludes with practical guidelines. All measurements are averaged over 20 runs.

\subsection{Latency Breakdown and the Staging Bottleneck}
\label{sec:latency}

\begin{figure*}[htbp]
\centering
\subfloat[]{\includegraphics[width=2.3in,height=1.7in]{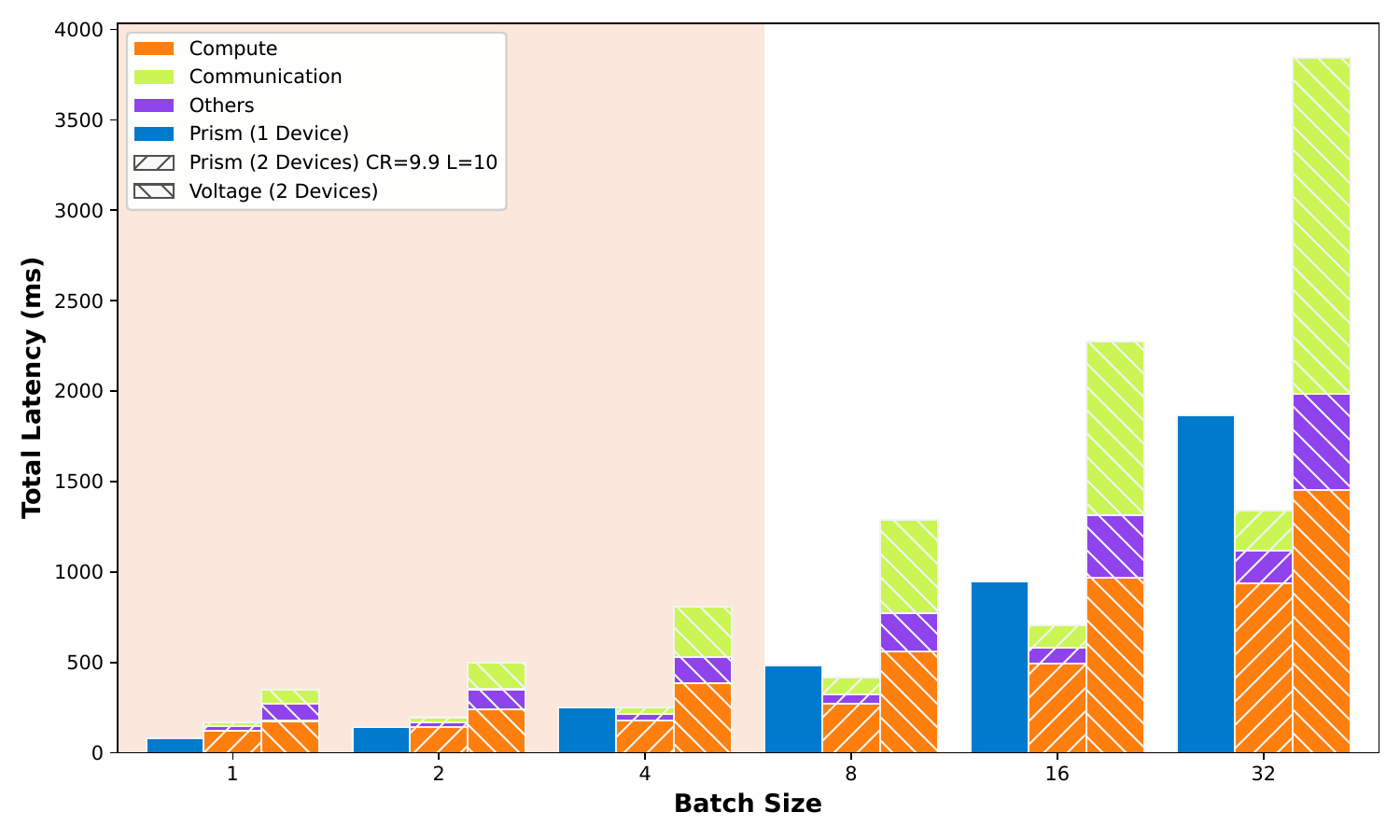}%
\label{fig:latency_breakdown}}
% \hfil
\subfloat[]{\includegraphics[width=2.3in,height=1.7in]{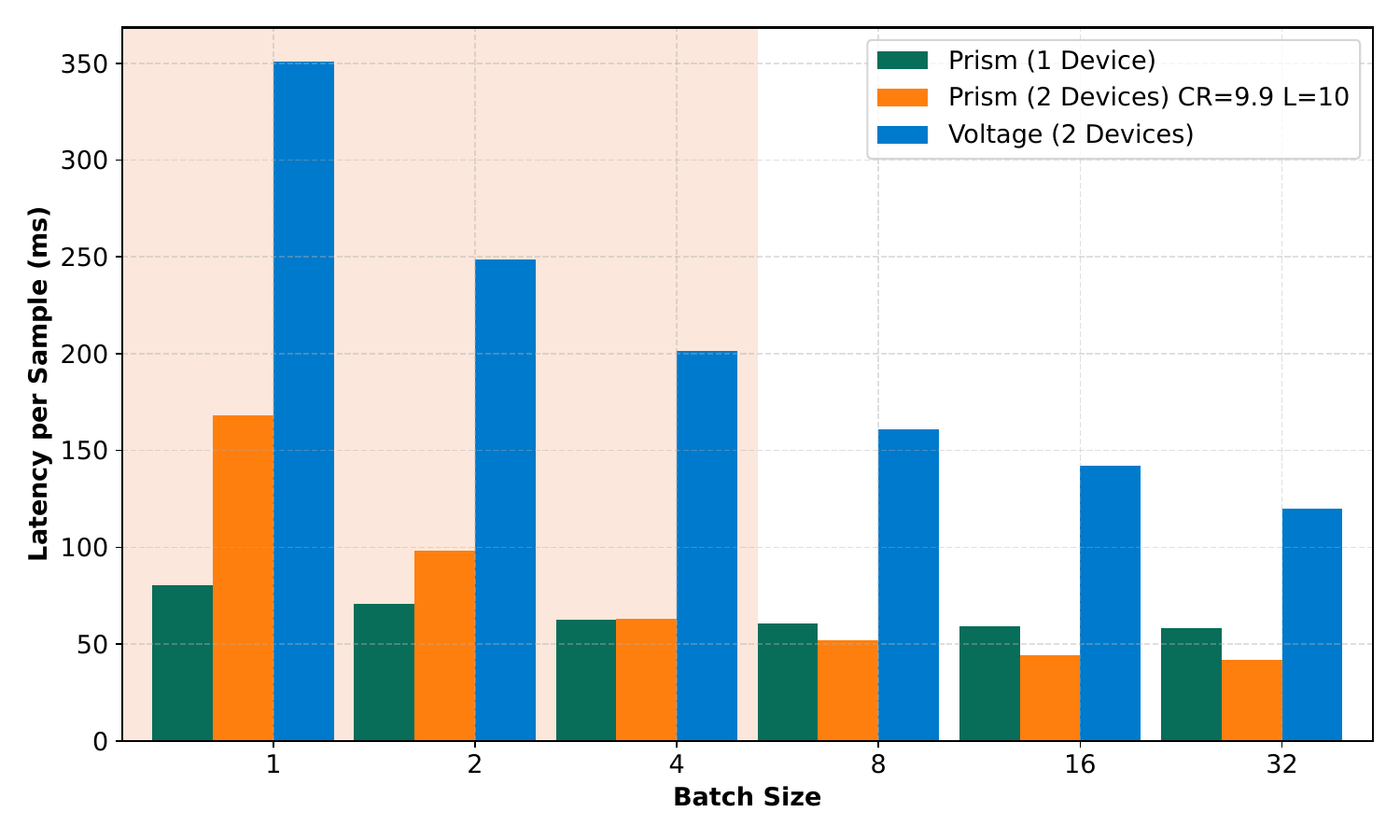}%
\label{fig:persamplelatency}}
% \hfil
\subfloat[]{\includegraphics[width=2.3in,height=1.7in]{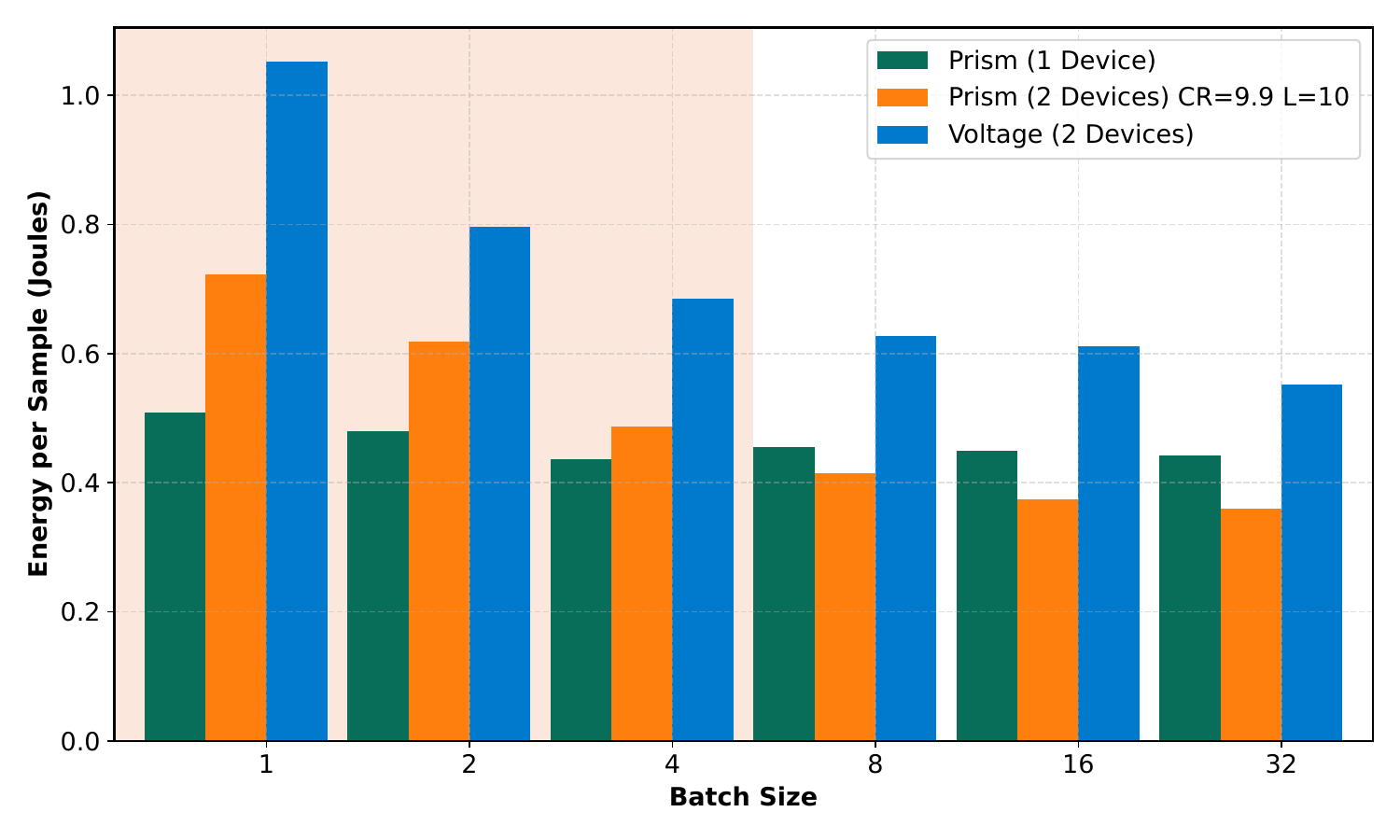}%
\label{fig:persampleenergy}}
\caption{Latency and energy scaling of distributed ViT inference on two Jetson Orin Nano devices.
(a) Total latency breakdown.
(b) Per-sample latency.
(c) Per-sample energy consumption.}
\label{fig_sim}
\end{figure*}

Fig.~\ref{fig:latency_breakdown} shows the latency breakdown across batch sizes at $\approx 400$~Mbps, with detailed numbers in Table~\ref{tab:latency}.
Three key observations emerge. 
% \begin{figure}[t]
%   \centering
%   \includegraphics[width=\linewidth]{images/total_latency_vs_batch_sizes.pdf}
%   \caption{Latency breakdown on two Jetson~Orin~Nano boards, showing dominant CPU–GPU staging for \voltage{} and single‑device selection regions.}
%   \label{fig:latency_breakdown}
% \end{figure}
% \begin{figure*}[t]
%   \centering

%   \begin{subfigure}[t]{0.32\textwidth}
%     \centering
%     \includegraphics[width=\linewidth]{images/total_latency_vs_batch_sizes.pdf}
%     \caption*{(a)}
%     \label{fig:latency_breakdown}
%   \end{subfigure}
%   \hfill
%   \begin{subfigure}[t]{0.32\textwidth}
%     \centering
%     \includegraphics[width=\linewidth]{images/latency_vs_batch_d2_prism.pdf}
%     \caption*{(b)}
%     \label{fig:persamplelatency}
%   \end{subfigure}
%   \hfill
%   \begin{subfigure}[t]{0.32\textwidth}
%     \centering
%     \includegraphics[width=\linewidth]{images/Energy_vs_batch_d2_prism.pdf}
%     \caption*{(c)}
%     \label{fig:persampleenergy}
%   \end{subfigure}

% \caption{Latency and energy scaling of distributed ViT inference on two Jetson Orin Nano devices.
% (a) Total latency breakdown.
% (b) Per-sample latency.
% (c) Per-sample energy consumption.}
%   \label{fig:batch_scaling}
% \end{figure*}

\begin{table}[t]
    \centering
    \caption{Latency breakdown across execution modes.}
    \label{tab:latency}
    \addtolength{\tabcolsep}{-4.2pt}
    % \footnotesize
    \begin{tabular}{m{1.2cm}m{0.3cm}m{0.5cm}m{0.5cm}m{0.5cm}ccccc}
        \toprule
        \multirow{2}{*}{\textbf{Model}} & \multirow{2}{*}{\textbf{$P$}} & \multirow{2}{*}{\textbf{$N$}} & \multirow{2}{*}{\textbf{$L$}} & \multirow{2}{*}{\textbf{$CR$}} & \multirow{2}{*}{\textbf{Batch}} & \multicolumn{4}{c}{\textbf{Latency (ms)}} \\
        \cmidrule(lr){7-10}
        & & & & & & \textbf{Comp.} & \textbf{Other} & \textbf{Comm.} & \textbf{Total} \\
        \midrule
        \multirow{6}{*}{\prism{}} & 1 & 197 & -- & -- & 1 & 80.6 & 0.0 & 0.0 & \textcolor{green}{\textbf{80.6}} \\
        & 1 & 197 & -- & -- & 2 & 141.3 & 0.0 & 0.0 & 141.3 \\
        & 1 & 197 & -- & -- & 4 & 249.8 & 0.0 & 0.0 & 249.8 \\
        & 1 & 197 & -- & -- & 8 & 485.0 & 0.0 & 0.0 & 485.0 \\
        & 1 & 197 & -- & -- & 16 & 946.0 & 0.0 & 0.0 & 946.0 \\
        & 1 & 197 & -- & -- & 32 & 1864.8 & 0.0 & 0.0 & 1864.8 \\
        \midrule
        \multirow{6}{*}{\prism{}} & 2 & 99 & 10 & 9.9 & 1 & 123.0 & \textcolor{green}{\textbf{26.5}} & 18.6 & 168.1 \\
        & 2 & 99 & 10 & 9.9 & 2 & 140.2 & 29.8 & 26.4 & 196.4 \\
        & 2 & 99 & 10 & 9.9 & 4 & 179.5 & 34.4 & 39.0 & 252.9 \\
        & 2 & 99 & 10 & 9.9 & 8 & 272.0 & 52.3 & 90.4 & \textcolor{blue}{\textbf{414.7}} \\
        & 2 & 99 & 10 & 9.9 & 16 & 494.0 & 86.7 & 124.0 & \textcolor{blue}{\textbf{704.7}} \\
        & 2 & 99 & 10 & 9.9 & 32 & 936.1 & 182.0 & 221.7 & \textcolor{blue}{\textbf{1339.8}} \\
        \midrule
        \multirow{6}{*}{\voltage{}} & 2 & 99 & 99 & 1 & 1 & 176.0 & \textcolor{red}{\textbf{94.0}} & 81.0 & 351.0 \\
        & 2 & 99 & 99 & 1 & 2 & 240.5 & 111.0 & 146.0 & 497.5 \\
        & 2 & 99 & 99 & 1 & 4 & 385.0 & 145.0 & 276.0 & 806.0 \\
        & 2 & 99 & 99 & 1 & 8 & 561.0 & 213.0 & 514.0 & 1288.0 \\
        & 2 & 99 & 99 & 1 & 16 & 970.0 & 344.0 & 960.5 & 2274.5 \\
        & 2 & 99 & 99 & 1 & 32 & 1454.0 & 533.0 & 1856.0 & 3843.0 \\
        \bottomrule
    \end{tabular}
\end{table}

\textbf{Staging dominates for \voltage{}.}
% The CPU--GPU I/O transfer (``Others'' in Fig.~\ref{fig:latency_breakdown}) is the critical latency bottleneck for \voltage{} at every batch size, scaling linearly with the volume of staged data.
% At batch~1, this staging cost alone reaches 94\,ms (Table~\ref{tab:latency}), already exceeding the total single-device latency of 80.6\,ms.
% This confirms that the \texttt{GLOO}-imposed memory staging negates the computational benefit of distributing the workload.
CPU–GPU I/O transfer (“Others” in Fig.~\ref{fig:latency_breakdown}) is the dominant latency bottleneck for \voltage{} at all batch sizes, scaling linearly with the staged data volume. At batch~1, staging alone reaches 94\,ms (Table~\ref{tab:latency}), already exceeding the single‑device latency of 80.6\,ms, confirming that \texttt{GLOO}-imposed staging negates the benefits of distributed execution.

\textbf{\prism{} suppresses staging cost.}
At CR~=~9.9, \prism{} exchanges only $L{=}10$ \sm{} vectors per partition per block instead of $N/P{=}98$ full token embeddings, reducing the staged tensor volume by approximately 90\%.
This brings the staging latency down to 26.5\,ms at batch~1 (Table~\ref{tab:latency}), shifting the bottleneck from communication back to computation.

\textbf{Adaptive crossover at batch~8.}
Below batch~8, even \prism{}'s reduced communication does not fully offset the fixed staging and synchronisation cost, so the profiling policy selects single-device execution (orange region in Fig.~\ref{fig:latency_breakdown}).
At batch~1, this yields 80.7\,ms---a 77\% reduction over \voltage{}'s 351\,ms.
From batch~8 onward, distributed \prism{} consistently outperforms single-device inference. 

% \begin{table}[!ht]
% \centering
% \caption{Computation and communication efficiency for ViT on CIFAR-10.}
% \label{tab:strategy_comparison}
% \begin{tabular}{>{\raggedright\arraybackslash}p{1.6cm}cm{1.0cm}m{1.0cm}m{0.3cm}m{1.0cm}m{1.0cm}}
% \toprule
% \multirow{2}{*}{\textbf{Strategy}} & \multirow{2}{*}{\textbf{P}} & \textbf{GFLOPs} & \textbf{Comp. Speed-up} & \multirow{2}{*}{\textbf{CR}} & \textbf{Comm. Speed-up} & \textbf{CIFAR-10} \\
% && \textbf{/device}&\textbf{\%}&&\textbf{\%}&Acc.\\
% \midrule
% No partition & 1 & 35.15 & - & - & - & 98.01 \\
% Voltage \cite{10631032} & 2 & 20.37 & 42.05 & - & - & 98.01 \\
% \midrule
% \multirow{3}{*}{\textbf{\prism{}}}& 2 & 17.54 & 50.11 & 9.90 & 89.90 & 95.64 \\
% & 2 & 17.86 & 49.20 & 4.95 & 79.80 & 96.84 \\
% & 2 & 18.18 & 48.29 & 3.30 & 69.70 & 97.06 \\
% \midrule
% \multirow{2}{*}{\shortstack{\textbf{\prism} \\ \textbf{(Finetuned)}}}  & 2 &  17.54 &  \textbf{50.11} & 9.90 &  \textbf{89.90} & \textbf{97.43}\\[2.2ex]
% \bottomrule
% \end{tabular}
% \end{table}

\begin{table}[!t]
\centering
\scriptsize
\setlength{\tabcolsep}{3pt}
\renewcommand{\arraystretch}{0.9}
\caption{Comparison of Computation and Communication Efficiency For ViT Model}
\label{tab:strategy_comparison}
\begin{tabular}{p{1.4cm} c p{1.3cm}   p{1.3cm}
  p{0.8cm}
  p{1cm}
  p{1cm}}
\toprule
\textbf{Strategy} & \textbf{$P$} & \textbf{GFLOPs /dev} &
\textbf{Comp. SU (\%)} & \textbf{$CR$} &
\textbf{Comm. SU (\%)} & \textbf{CIFAR-10 Acc.} \\
\midrule
No partition & 1 & 35.15 & - & - & - & 98.01 \\
Voltage \cite{10631032} & 2 & 20.37 & 42.05 & - & - & 98.01 \\
\midrule
\multirow{3}{*}{\textbf{\prism}} 
 & 2 & 17.54 & 50.11 & 9.90 & 89.90 & 95.64 \\
 & 2 & 17.86 & 49.20 & 4.95 & 79.80 & 96.84 \\
 & 2 & 18.18 & 48.29 & 3.30 & 69.70 & 97.06 \\
\midrule
\textbf{\prism}  

 & 2 & 17.54 & \textcolor{blue}{\textbf{50.11}} &
 9.90 & \textcolor{blue}{\textbf{89.90}} &
 \textcolor{blue}{\textbf{97.43}} \\
  \textbf{(Finetuned)} & & & & & & \\
\bottomrule
\end{tabular}
\end{table}

Table~\ref{tab:main_results} quantifies the overall gains.
\prism{} achieves 65.1\%--77.0\% latency reduction and 34.1\%--51.8\% energy reduction over \voltage{} across all tested batch sizes.
These gains come with a minor accuracy trade-off: as shown in Table~\ref{tab:strategy_comparison}, the CIFAR-10 accuracy at CR~=~9.9 drops from 98.01\% to 95.64\%, which recovers to 97.43\% after fine-tuning.
Although lower compression rates (CR~=~3.3) retain 97.06\% accuracy without fine-tuning, it slightly compromises communication speed‑up (i.e. 69.7\% vs.\ 89.9\%).

\begin{table}[t]
\caption{\prism{} vs.\ \voltage{} on Jetson~Orin~Nano boards (ViT, CIFAR‑10, CR~=~9.9, 400\,Mbps). Orange rows mark single‑device execution.}
  \label{tab:main_results}
  \centering
  \small
  \setlength{\tabcolsep}{3.8pt}
  \begin{tabular}{c rr r rr r}
    \toprule
    \multirow{2}{*}{\textbf{Batch}} &
    \multicolumn{3}{c}{\textbf{Latency (ms)}} &
    \multicolumn{3}{c}{\textbf{Energy (J)}} \\
    \cmidrule(lr){2-4}\cmidrule(lr){5-7}
    & \voltage{} & \prism{} & \textbf{Gain}
    & \voltage{} & \prism{} & \textbf{Gain} \\
    \midrule
    \rowcolor{orange!30}
    1  & 351.0  & 80.7   & 77.0\% & 1.05  & 0.51  & 51.8\% \\
    \rowcolor{orange!30}
    2  & 497.5  & 141.3  & 71.6\% & 1.59  & 0.96  & 39.6\% \\
    \rowcolor{orange!30}
    4  & 806.0  & 249.8  & 69.0\% & 2.74  & 1.75  & 36.2\% \\
    8  & 1288.0 & 414.7  & 67.8\% & 5.02  & 3.31  & 34.1\% \\
    16 & 2274.5 & 704.7  & 69.0\% & 9.78  & 5.98  & 38.8\% \\
    32 & 3843.0 & 1339.8 & 65.1\% & 17.67 & 11.52 & 34.8\% \\
    \bottomrule
  \end{tabular}
\end{table}

\begin{figure*}
    \centering
    \includegraphics[width=\textwidth]{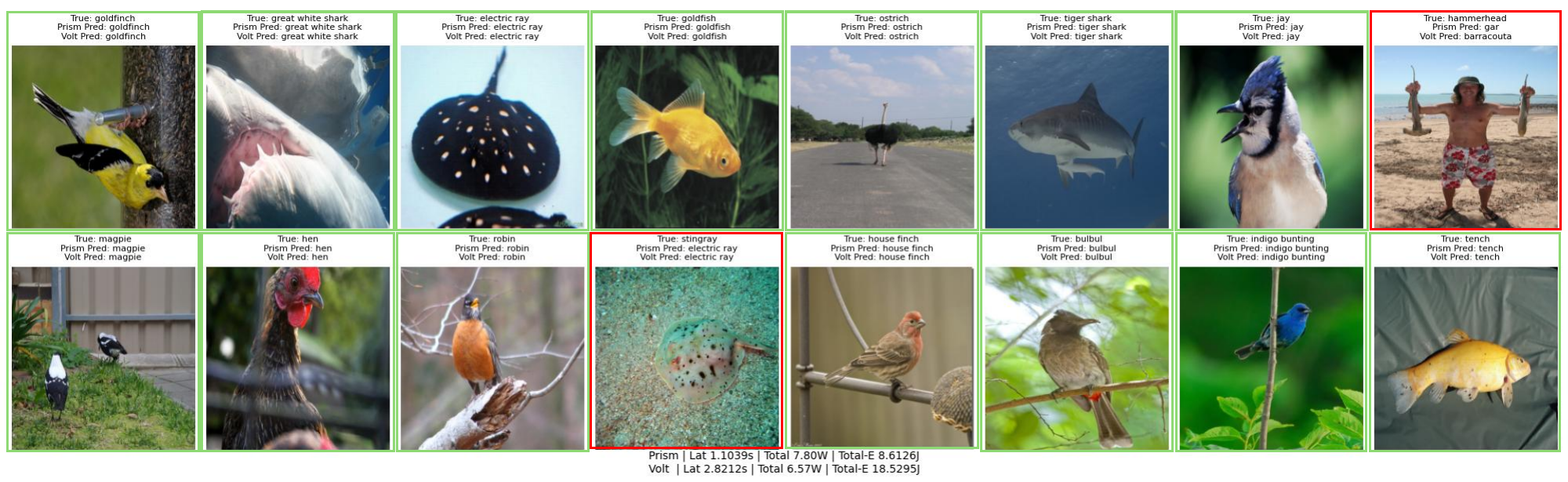}
    \caption{Batch-based inference comparison between \prism{} and \voltage{} on 16 CIFAR‑10 images, showing predicted and ground‑truth labels with per‑batch latency and energy.}
    \label{fig:batch_predictions}
\end{figure*}

\subsection{Per-Sample Latency and Energy}
\label{sec:per_sample}

Fig.~\ref{fig:persamplelatency} and Fig.~\ref{fig:persampleenergy} present per-sample latency and energy across batch sizes.

\textbf{Latency.}
\voltage{}'s per-sample latency exceeds single-device inference at every batch size, confirming that \texttt{GLOO} staging overhead is never amortized.
\prism{} combined with the profiling-driven adaptive policy, it reduces per-sample latency as batch size grows because the communication cost is spread over more samples.
At batch~32, \prism{} achieves approximately 42\,ms per sample, compared to 120\,ms for \voltage{} and 58\,ms for single-device execution.

\textbf{Energy.}
Per-sample energy closely tracks latency, since the boards operate at a fixed power mode.
\prism{} saves 51.8\% energy at batch~1 and 34.8\% at batch~32 relative to \voltage{} (Table~\ref{tab:main_results}).
From batch~8 onward, \prism{}'s per-sample energy closely approaches the single-device lower bound, indicating that overheads become smaller once communication volume is compressed.

\begin{figure}[t]
  \centering
  \includegraphics[width=\columnwidth]{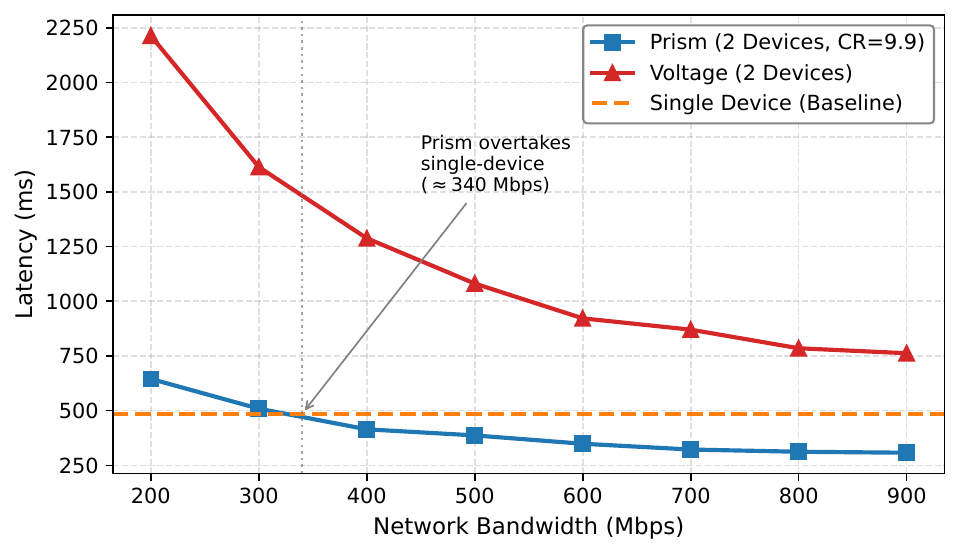}
  \caption{Per-sample latency vs.\ network bandwidth at $B{=}8$ on two Jetson Orin Nano boards (ViT, CIFAR-10).}
  \label{fig:bw_latency}
\end{figure}

\textbf{Bandwidth sensitivity.}
Fig.~\ref{fig:bw_latency} sweeps network bandwidth at $B{=}8$.
\voltage{} remains bandwidth-bound across 200--900~Mbps, while \prism{} crosses the single-device baseline near 340~Mbps and continues to improve above it, confirming bandwidth as a meaningful parameter of the profiling.

\textbf{Limitations.}
Despite these gains, distributed \prism{} does not reach the theoretical ideal of half the single-device latency, residual \texttt{GLOO} staging and synchronization costs remain non-negligible even with compressed communication.
At small batch sizes (1--4), these fixed costs dominate and the profiling policy falls back to single-device execution.
This matters for edge deployments where real-time applications often operate at $B \in [1,4]$; however, the crossover also depends on sequence length $N$, so workloads with large $N$ (e.g., LLM inference) can benefit from distributed execution even at $B{=}1$.
The decision to distribute therefore depends on both batch size and sequence length, motivating profiling-driven adaptation.

% \subsection{Bandwidth Sensitivity}
% \label{sec:bandwidth}

% \begin{table}[t]
% \centering
% \caption{Measured data rates for available connections.}
% \label{tab:datarates}
% \begin{tabular}{l l c}
% \toprule
% \makecell{\textbf{Connection}\\\textbf{Type}} &
% \makecell{\textbf{Band /}\\\textbf{Frequency}} &
% \makecell{\textbf{Measured}\\\textbf{Throughput}} \\
% \midrule
% Ethernet (LAN) & ---        & $\sim$850~Mbps \\
% Wi-Fi~6        & 6~GHz      & $\sim$900~Mbps \\
% Wi-Fi~5 (Upper)& 5~GHz$_2$  & $\sim$760~Mbps \\
% Wi-Fi~5 (Lower)& 5~GHz$_1$  & $\sim$660~Mbps \\
% Wi-Fi~2.4~GHz  & 2.45~GHz   & $\sim$210~Mbps \\
% \bottomrule
% \end{tabular}
% \end{table}
% Table~\ref{tab:datarates} reports the measured throughput across connection types on our testbed, ranging from approximately 210~Mbps (2.4~GHz) to 900~Mbps (WiFi~6 at 6~GHz).

% The latency breakdown in Section~\ref{sec:latency} reveals that \prism{}'s dominant overhead is CPU--GPU staging, which depends on tensor size and is largely independent of network bandwidth.
% \voltage{}, by contrast, is strongly bandwidth-sensitive: its large communication volume causes both network transfer time and staging time to grow substantially as throughput drops from 900 to 200~Mbps. This simplifies the profiling sweep and makes \prism{} well suited for variable-bandwidth environments such as real-world WiFi-connected edge clusters.

\subsection{Qualitative Inference Validation}
\label{sec:qualitative}

Beyond efficiency metrics, we verify that \prism{}'s compression preserves prediction correctness.
Fig.~\ref{fig:batch_predictions} shows a batch of 16 CIFAR-10 images processed by both \prism{} and \voltage{} on two Jetson Orin Nano boards.
Both methods produce identical predicted labels across all 16~samples.
Two images are misclassified by both methods; single-device inference produces the same errors.
These failures stem from visually ambiguous content (cluttered backgrounds) and are attributable to the ViT model itself, not to the \sm{} approximation or the partitioning scheme.

In terms of end-to-end efficiency, \prism{} completes this batch in 1.10\,s with 8.61\,J, compared to 2.30\,s and 18.50\,J for \voltage{}.
While this represents a substantial improvement, the distributed configuration still incurs non-negligible staging overhead compared to ideal scaling, reinforcing that \prism{} mitigates (rather than resolves) the fundamental \texttt{GLOO} staging bottleneck.

\subsection{Profiling Instrumentation}
\label{sec:profiling_instrumentation}

\begin{figure}[!htbp]
    \centering
    \includegraphics[width=\columnwidth,height=0.35\textheight,
        keepaspectratio,
        ]{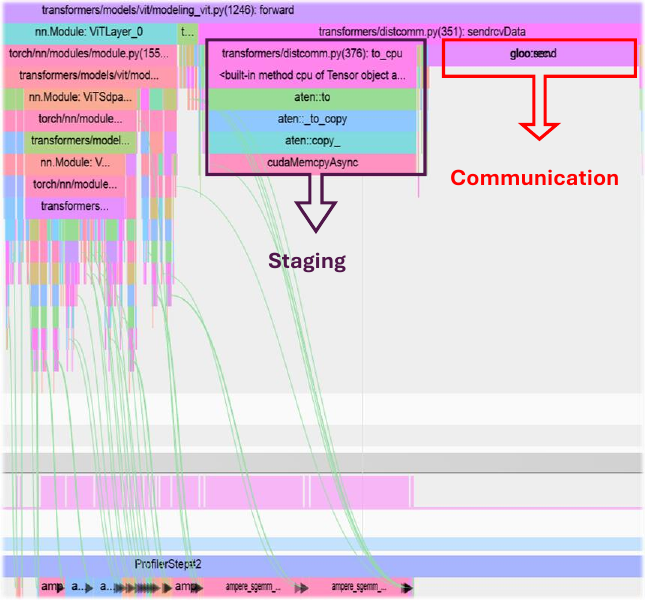}
    \caption{PyTorch Profiler trace of distributed \prism{} inference on a Jetson~Orin~Nano, illustrating computation, \texttt{GLOO} communication, and CPU–GPU staging.}
    \label{fig:profiler_trace}
\end{figure}

Fig.~\ref{fig:profiler_trace} shows a PyTorch Profiler trace captured during distributed inference.
The trace reveals three distinct phases within each Transformer block: GPU computation kernels, \texttt{GLOO} collective communication calls, and CPU--GPU memory staging operations.
This confirms the three-way decomposition used in Table~\ref{tab:latency} and validates that the offline profiling mechanism (Section~\ref{sec:adaptive}) can accurately attribute latency to each component when populating the performance map.

\subsection{Summary of Profiling Insights}
\label{sec:insights}
Our prototype experiments yield two concrete guidelines for practitioners.

\textbf{Always assume GLOO on Jetson; design for staging.}
Any distributed inference system on Jetson-class hardware must consider CPU--GPU staging bottleneck.
Full-tensor exchange methods for transformer models will likely degrade performance regardless of bandwidth.
While \prism{} reduces the staging volume by up to 90\%, the residual overhead is not eliminated and remains a limiting factor at small batch sizes.

\textbf{Profile; do not estimate.}
The crossover batch size at which distributed execution becomes beneficial cannot be estimated accurately through  modeling FLOPs analytically.
A one-time profiling sweep of ${\sim}200$ inference passes is sufficient to populate the performance map.

\section{Conclusion}
\label{sec:conclusion}
We presented a hardware prototype study of distributed Transformer inference on a set of NVIDIA Jetson Orin Nano boards connected over WiFi. Our central finding is that the \texttt{GLOO} backend imposes a CPU--GPU staging overhead that renders methods relying on full-tensor exchange like \voltage{} slower than single-device inference at every tested batch size. \prism{}'s \sm{} compression reduces this staging volume by up to 90\%, delivering 65\%--77\% latency and 34\%--52\% energy reductions over \voltage{}, while residual staging costs are not completely eliminated and remain a limiting factor at small batch sizes. A profiling-driven adaptive policy addresses this gap by querying an offline performance map at runtime to select single-device or distributed execution as appropriate. We validated these findings and confirmed that \prism{}'s compression preserves the model's prediction behavior. We believe that this developed testbed, profiling methodology, and rigorous characterization of both the benefits and remaining limitations of \texttt{GLOO}-based distributed inference provide a practical foundation for deploying foundation models at the edge and motivate future research on improved communication backends compatible with integrated GPUs.
%% ==============================================================
\begin{acks}
This research was supported by the PANDORA project, funded
by the European Union's Horizon Europe Framework Programme
under Grant Agreement No~101135775, and by NordForsk Nordic
University Cooperation on Edge Intelligence (Grant
No.~168043).
\end{acks}
%% ==============================================================
\bibliographystyle{ACM-Reference-Format}
\bibliography{base}
\end{document}